\documentclass[12pt]{article}
\usepackage[margin=2.5cm]{geometry}

\usepackage{amsthm,amsmath}
\usepackage{amsmath}
\usepackage{amssymb}
\usepackage{amsbsy}
\usepackage{amsthm}
\usepackage{mathtools}
\usepackage{graphicx}
\usepackage{multirow}
\usepackage{enumerate}
\usepackage{xcolor}
\usepackage{tikz}
\usepackage{rotating}
\usepackage{url}
\usepackage{pbox}
\usepackage{titlesec}
\usepackage{hyperref}
\usepackage{natbib}

\titlespacing*{\section}{0pt}{0.75\baselineskip}{0.5\baselineskip}

\title{\textbf{AMICI: High-Performance Sensitivity Analysis for Large Ordinary Differential Equation Models}}

\author{
Fabian Fröhlich\,$^{1,*}$,
Daniel Weindl\,$^{2}$, 
Yannik Schälte\,$^{2, 3}$,\\
Dilan Pathirana\,$^{4}$, 
\L{}ukasz Paszkowski\,$^{5}$,
Glenn Terje Lines\,$^{5}$,\\
Paul Stapor\,$^{2, 3}$ and
Jan Hasenauer\,$^{2,3,4,*}$ 
}

\date{December 2020}

\begin{document}

\maketitle
{\parindent0pt
\small
$^1$Department of Systems Biology, Harvard Medical School, Boston, MA 02115, USA,\\
$^2$Institute of Computational Biology, Helmholtz Zentrum München -- German Research Center for Environmental Health, 85764 Neuherberg, Germany,\\
$^3$Center for Mathematics, Technische Universität München, 85748 Garching, Germany,\\
$^4$Faculty of Mathematics and Natural Sciences, University of Bonn, 53113 Bonn, Germany, and \\
$^5$Simula Research, 1325 Lysaker, Norway.\\
$^*$To whom correspondence should be addressed.
}

\begin{abstract}
\noindent\textbf{Summary:} Ordinary differential equation models facilitate the  understanding of cellular signal transduction and other biological processes. However, for large and comprehensive models, the computational cost of simulating or calibrating can be limiting. AMICI is a modular toolbox implemented in C++/Python/MATLAB that provides efficient simulation and sensitivity analysis routines tailored for scalable, gradient-based parameter estimation and uncertainty quantification. \\
\noindent\textbf{Availability:} AMICI is published under the permissive BSD-3-Clause license with source code publicly available on \url{https://github.com/AMICI-dev/AMICI}. Citeable releases are archived on Zenodo.  \\
\noindent\textbf{Contact:}
\href{jan.hasenauer@uni-bonn.de}{jan.hasenauer@uni-bonn.de}, \href{fabian\_froehlich@hms.harvard.edu}{fabian\_froehlich@hms.harvard.edu}
\\
\noindent\textbf{Supplementary information:} Supplementary information is available at arXiv
online.
\end{abstract}

\section{Introduction}

Ordinary Differential Equation (ODE) models are widely used in systems biology, for example, to elucidate dynamic processes and to predict response to perturbations. 
Model parameters have to be inferred from data, which can be computationally intensive as thousands of simulations may be required. This is challenging for large models, with many state variables and parameters, where simulations take seconds to minutes \citep{FroehlichKes2018}. Inference can benefit from accurate sensitivities \citep{VillaverdeFro2018}, i.e., derivatives of model outputs with respect to model parameters. Accurate sensitivities can be computed using forward, adjoint or steady-state approaches, but benefit from symbolic derivatives of model equations, which are labor-intensive and error-prone to compute manually. 

The recent surge of genome-scale perturbation data as well as respective comprehensive models has increased the demand for methods for scalable sensitivity computation for ODE models. To address this demand, we introduce AMICI, a high-performance simulation and sensitivity analysis library. AMICI is implemented in C++ and Python, and provides model import from widely used formats such as the Systems Biology Markup Language (SBML) \citep{HuckaFin2003}, BioNetGen Language (BNGL) \citep{HarrisHog2016} and Kappa \citep{BoutillierMaa2018}, and generates high-performance-computing (HPC) ready modules.
These modules provide model-specific simulation and sensitivity computation routines, which can be accessed from Python, C++, and MATLAB. For parameter estimation problems specified in PEtab \citep{schmiester2020petab}, AMICI can evaluate the objective function and its gradient.

\section{Methods}

For high simulation performance, AMICI reads models from high-level formats, then translates the model and symbolically derived expressions to C++ code. As symbolic processing can be computationally intensive, AMICI symbolically only computes partial derivatives, total derivatives are computed through (sparse) matrix multiplication and addition at runtime. This expedites model compilation and simulation.

To simulate stiff, large models, efficient linear solvers are crucial. AMICI features several direct dense, direct sparse and implicit solvers. For sensitivity analysis, AMICI implements forward, adjoint \citep{FroehlichKal2017}, and steady-state \citep{LinesPas2019} methods and combinations thereof (see Supplementary Information).

\section{Implementation}

The AMICI library is implemented in C++14 and relies on SUNDIALS \citep{HindmarshBro2005} and SuiteSparse \citep{DavisPal2010} for simulation and sensitivity computation. For matrix multiplication, a CBLAS-compatible library is used (e.g.\ ATLAS, OpenBLAS, Intel MKL).

\begin{figure}[t]
\centering
\includegraphics[width=0.7\textwidth]{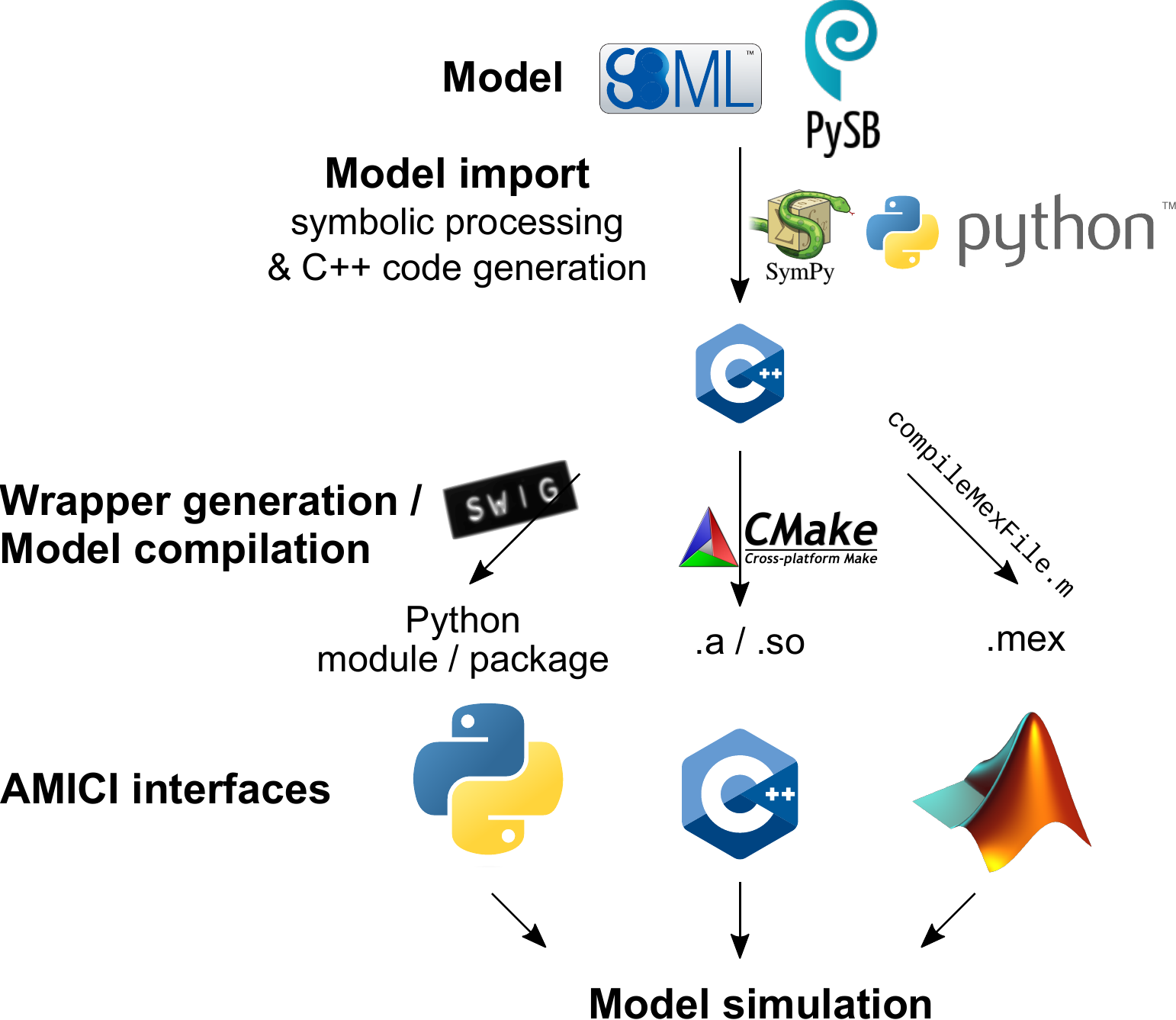}

\caption{Outline of model compilation in AMICI. Models, specified in SBML or PySB, are imported in Python. This generates efficient C++ code for model equations and simulation bindings for Python, C++ and MATLAB.}
\label{fig:workflow}
\end{figure}

Model import is implemented in Python and supports several widely used formats (Figure~\ref{fig:workflow}). SBML import is implemented using libSBML \citep{BornsteinKea2008}, Kappa and BNGL import is implemented via PySB \citep{LopezMuh2013}. Symbolic processing is performed using SymPy \citep{MeurerSmi2017}. Compilation has been tested with GCC, Clang, Intel and MinGW on Linux, MacOS and Windows platforms. For interoperability, AMICI provides CMake files and Dockerfiles.

AMICI ships with a MATLAB interface and implements a Python interface using SWIG, which can be extended to other languages. Simulation and compilation are highly configurable with the API documented on \href{https://amici.rtfd.org/}{Read the Docs} and in example notebooks. For HPC-readiness, AMICI implements OpenMP parallelization over experimental conditions and is thread-safe.

Simulation and sensitivity analysis rely on intricate theory that is complex, error-prone to implement. Accordingly, we implemented an extensive continuous integration test pipeline. Simulation results are verified for the SBML semantic test suite, where AMICI passes 862 out of 1780 tests (appropriate error messages to indicate unsupported features for remaining tests). Simulation results are compared against PySB simulations for 17 BNGL validation models and examples, where AMICI passes all tests. Sensitivity results are checked by regression tests covering, e.g.\, forward, adjoint and steady-state sensitivities. Performance tests check computation time for model import, simulation and sensitivity analysis of a large model. Unit and documentation tests, static code analysis, and memory leak checks are included, and code review is enforced for all contributions.

\section{Discussion}

There is a rich ecosystem of tools for model simulation. To avoid duplication, we ensured good interoperability with other tools: AMICI does not provide a model development environment, but permits model import from standard formats. Similarly, AMICI is not part of an integrated parameter estimation framework, but features a flexible, well-documented API. Currently, several parameter estimation tools can interface AMICI, including pyPESTO \citep{SchaelteFro2020} and parPE \citep{SchmiesterSch2019}. This modular design aims at researchers developing new methods or tools for parameter estimation that would benefit from state-of-the-art simulation and sensitivity computation routines. Specifically, adjoint and steady-state sensitivity analysis as well as sparse linear solvers are currently only supported in a small, disparate set of tools, which highlights the unique capabilities of AMICI.

AMICI has been in development since 2015 and has so far been used in at least 50 publications, and is continuously being developed by 4 core contributors at 3 different institutions. In the future, we plan to improve SBML support and extend interoperability with other tools.

\section*{Funding}

This work was supported by the European Union's Horizon 2020 research and innovation program (CanPathPro; Grant no. 686282; J.H., D.W., P.S., \L.P., G.T.L., F.F.), the Federal Ministry of Education and Research of Germany (Grant no.\ 01ZX1916A; D.W. \& 01ZX1705A; J.H., 
\& Grant. no.\ 031L0159C; J.H.), 
the German Research Foundation (Grant no. HA7376/1-1; Y.S., Germany's Excellence Strategy -- EXC-2047/1 -- 390685813; D.P.), the Human Frontier Science Program (Grant no.\ LT000259/2019-L1; F.F.), the National Institute of Health (Grant no.\ U54-CA225088; F.F.), and 
the Federal Ministry of Economic Affairs and Energy (Grant no. 16KN074236; D.P.).

\bibliographystyle{apalike}
\bibliography{document}
\end{document}